\documentclass[10pt,conference]{IEEEtran}
\usepackage{soul}
\usepackage{amsmath}
\usepackage{}
\usepackage{overpic}
\usepackage{verbatim} 
\usepackage{pifont}
\usepackage{times,amsmath,color,amssymb,graphicx, epstopdf, epsfig,cite,psfrag,subfigure,algorithm}
\usepackage[noend]{algpseudocode}
\usepackage{enumerate}
\usepackage{xcolor}

\usepackage{color}
\IEEEoverridecommandlockouts

\begin{document}
	
	
	%
	\title{
Online Network Slicing for Real Time Applications in Large-scale Satellite Networks
	\thanks{This work is supported by the Natural Science Foundation of China (61931017). The corresponding author is Hongyan Li.
}	
}
\author{\IEEEauthorblockN{Binquan Guo$^*$$^\dagger$, Hongyan~Li$^*$, Zhou Zhang$^\dagger$, Ye Yan$^\dagger$}\\
	\IEEEauthorblockA{$^*$State Key Laboratory of Integrated Service Networks, Xidian University, Xi'an P. R. China\\
		$^\dagger$Tianjin Artificial Intelligence Innovation Center (TAIIC), Tianjin, P. R. China\\
		Email: bqguo@stu.xidian.edu.cn, hyli@xidian.edu.cn, zt.sy1986@163.com, yanye1971@sohu.com}}
	\maketitle
	
	
\begin{abstract}
In this work, we investigate resource allocation strategy for real time communication (RTC) over satellite networks with virtual network functions. Enhanced by inter-satellite links (ISLs), in-orbit computing and network virtualization technologies, large-scale satellite networks promise global coverage at low-latency and high-bandwidth for RTC applications with diversified functions. However, realizing RTC with specific function requirements using intermittent ISLs, requires efficient routing methods with fast response times. We identify that such a routing problem over time-varying graph can be formulated as an integer linear programming problem. The branch and bound method incurs $\mathcal{O}(|\mathcal{L}^{\tau}| \cdot (3 |\mathcal{V}^{\tau}| + |\mathcal{L}^{\tau}|)^{|\mathcal{L}^{\tau}|})$ time complexity, where $|\mathcal{V}^{\tau}|$ is the number of nodes, and $|\mathcal{L}^{\tau}|$ is the number of links during time interval ${\tau}$. By adopting a k-shortest path-based algorithm, the theoretical worst case complexity becomes $O(|\mathcal{V}^{\tau}|! \cdot |\mathcal{V}^{\tau}|^3)$. Although it runs fast in most cases, its solution can be sub-optimal and may not be found, resulting in compromised acceptance ratio in practice. To overcome this, we further design a graph-based algorithm by exploiting the special structure of the solution space, which can obtain the optimal solution in polynomial time with a computational complexity of $\mathcal{O}(3|\mathcal{L}^{\tau}| + (2\log{|\mathcal{V}^{\tau}|}+1) |\mathcal{V}^{\tau}|)$. Simulations conducted on starlink constellation with thousands of satellites corroborate the effectiveness of the proposed algorithm.

	\end{abstract}
	

	%
	\IEEEpeerreviewmaketitle

	\section{Introduction}
In recent years, commercial enterprises such as SpaceX, Amazon and Oneweb are building up mega-constellations with hundreds or thousands of low earth orbit (LEO)  satellites to provide global coverage at low-latency and high-bandwidth. As reported by ITU \cite{ITU2021}, nearly $2.9$ billion people (around $37\%$ of the world's population) still do not have access to Internet. More than $70\%$ of the surface of the earth has no terrestrial network coverage (e.g., in ocean, dessert and rain forest areas). Satellite networks (SNs) can not only supplement current terrestrial networks in areas short of Internet infrastructure, but also help provide data communication services for areas where the terrestrial networks are deployed but highly loaded.
Enhanced by inter-satellite links (ISLs) and on-board computing resources, SNs have potential to offer real time applications for ground users with a latency in tens of milliseconds \cite{lai2022spacertc}. 

Traditional satellites are customized for particular tasks, without cooperation and resulting in resource under utilization and high operation cost \cite{sheng2017toward}. 
It is not flexible for customized SNs to update software or develop new functions, which brings great challenges to rapid adoption of advanced strategies to improve network performance. To address these challenges, virtualization technologies, including software defined networking (SDN) and network function virtualization (NFV), have been introduced into the SNs. In particular, NFV decouples network functions from the dedicated hardwares, thereby, network functions can be virtualized into software components or abstracted as containers, which are referred to as virtual functions (VFs).
In this way, VFs can be flexibly deployed on different satellites to establish customized virtual networks for diversified requested services. Each requested service can specify its targeted VF and quality of service (QoS) requirements such as transmission capacity and end to end delay. 
Both the academia and industry efforts have been made on the feasibility of using NFV in the SNs. 
In \cite{xu2018software}, NFV is exploited to facilitate the incorporation of new applications in the SNs. The benefits of applying NFV into the SNs are verified through various use cases in \cite{bertaux2015software}. 
The implementation of NFV in SNs is investigated in \cite{zhou2019bidirectional}. 
The key technologies of NFV have already been verified by a couple of satellites in space, such as Tianzhi 1 and Eutelsat Quantum.
However, research on routing for real time applications over SN with VFs is absent, which is the main focus of our work. 

Despite the flexibility and benefits introduced by NFV, enabling real time communications in SNs with NFV still faces many challenges. The multimedia-based real time applications, such as video-conferencing, Internet telephony and interactive VR/AR applications, are delay sensitive in nature. To provision a requested real time service, the routing strategy must have fast response time and satisfy both the QoS and network function requirements under the time-varying topologies.
Therefore, the routing strategy becomes more complex than traditional ones designed for terrestrial networks. 
Additionally, there are few efforts to investigate routing in SNs with NFV. In \cite{wang2020sfc, jia2021vnf, yang2021maximum}, the optimization of NFV deployment and routing strategy in the SNs was investigated. However, their \textit{store-wait-forward} data transmission mechanism is designed for delay tolerant tasks, which can incur unacceptable latency in the order of minutes or even hours (especially in large scale SNs) and is not applicable for delay sensitive services.

To fill this gap, we investigate the routing strategy for real time applications in the SNs with NFVs. We identify that such a routing problem over the time-varying graph can be formulated as an integer linear programming problem. The branch and bound method for solving it incurs $\mathcal{O}(|\mathcal{L}^{\tau}| \cdot (3 |\mathcal{V}^{\tau}| + |\mathcal{L}^{\tau}|)^{|\mathcal{L}^{\tau}|})$ time complexity, where $|\mathcal{V}|^{\tau}$ is the number of nodes, and $|\mathcal{L}^{\tau}|$ is the number of links in the snapshot graph within time interval $\tau$. By adopting a k-shortest path-based algorithm, the worst-case time complexity becomes $O(|\mathcal{V}^{\tau}|! \cdot |\mathcal{V}^{\tau}|^3)$, but is fast enough in practical cases. However, its solution can be suboptimal, which may result in compromised acceptance ratio. To overcome this, we design an alternative algorithm by exploiting the special structures of the solution space, which can obtain the optimal solution in polynomial time with a low computation complexity of $\mathcal{O}(3|\mathcal{L}^{\tau}| + (2\log{|\mathcal{V}^{\tau}|}+1) |\mathcal{V}^{\tau}|)$ and perform more stable in large scale networks. Simulations conducted on starlink constellation with thousands of satellites corroborate the effectiveness of the proposed algorithm. 

	\section{System Model and Problem Formulation}
			\subsection{Satellite network scenario}

	We consider one typical SN composed of satellites and ground terminals (GTs), which are denoted by $\mathcal{S} = \{\mathbb{S}_1, ..., \mathbb{S}_P\}$ and $\mathcal{O} = \{\mathbb{O}_1, ..., \mathbb{O}_K\}$, respectively. Here $P$ and $K$ are the numbers of elements in $\mathcal{S}$ and $\mathcal{O}$, respectively. Each satellite is equipped with computing and communication hardwares. The computing resources are defined into different types of VFs (such as object recognition, data mining, information encryption and signal processing) by taking advantage of SDN, NFV and micro-services techniques, which guarantees that different types of services can be provisioned in the same SN with flexibility. ISLs are supported in the SN, which are intermittent and predictable with satellites' movement.

	Denote the set of functions supported in the SN as $\mathcal{F} = \{f_1, f_2, ..., f_N\}$, where $N = |\mathcal{F}|$. We assume different numbers and types of functions have been deployed on different satellites in the SN. Specifically, each satellite can provide one or multiple functions, which is different from our previous work \cite{yang2021maximum} assuming only one function is supported per satellite. To prevent the functions of a single satellite from being called by excessive applications, the maximum allowable call number of a function $f_l \in \mathcal{F}$ in a satellite $\mathbb{S}_i \in \mathcal{V}$ is set as $w^{f_l}_{\mathbb{S}_i} $, where $w^{f_l}_{\mathbb{S}_i} \in \{0, 1, 2, 3,...\}$. Specifically, if $w^{f_l}_{\mathbb{S}_i} > 0$, satellite $\mathbb{S}_i$ can handle applications requiring $f_l$ function; if $w^{f_l}_{\mathbb{S}_i} = 0$, no more applications with function $f_l$ can be provisioned. 
	
	Given a time horizon $\mathcal{T}$, we use $\mathcal{A} = \{\mathbb{O}_s, \mathbb{O}_d, f_a, C_a, D_a \}$ to denote the real time communication application requiring function $f_a$, where $\mathbb{O}_s$ denotes the source GT, $\mathbb{O}_d$ denotes the destination GT, $C_a$ is the required transmission bandwidth, and $D_a$ is the maximum acceptable end to end delay from $\mathbb{O}_s$ to $\mathbb{O}_d$. In this work, we assume each application requires one function, and the corresponding data processing is conducted on one single satellite with no split. In other words, an end to end feasible path $p$ for $\mathcal{A}$ must contain at least one satellite providing function $f_a$ for data flow processing, while the other satellites within the path $p$ only relay data without providing functions.
    Besides, the transmission bandwidth and end to end delay requirements are also specified by the user applications, however, how long the application will last is unknown.

	\begin{figure}[t!]
		\centering
				\includegraphics[width=70mm]{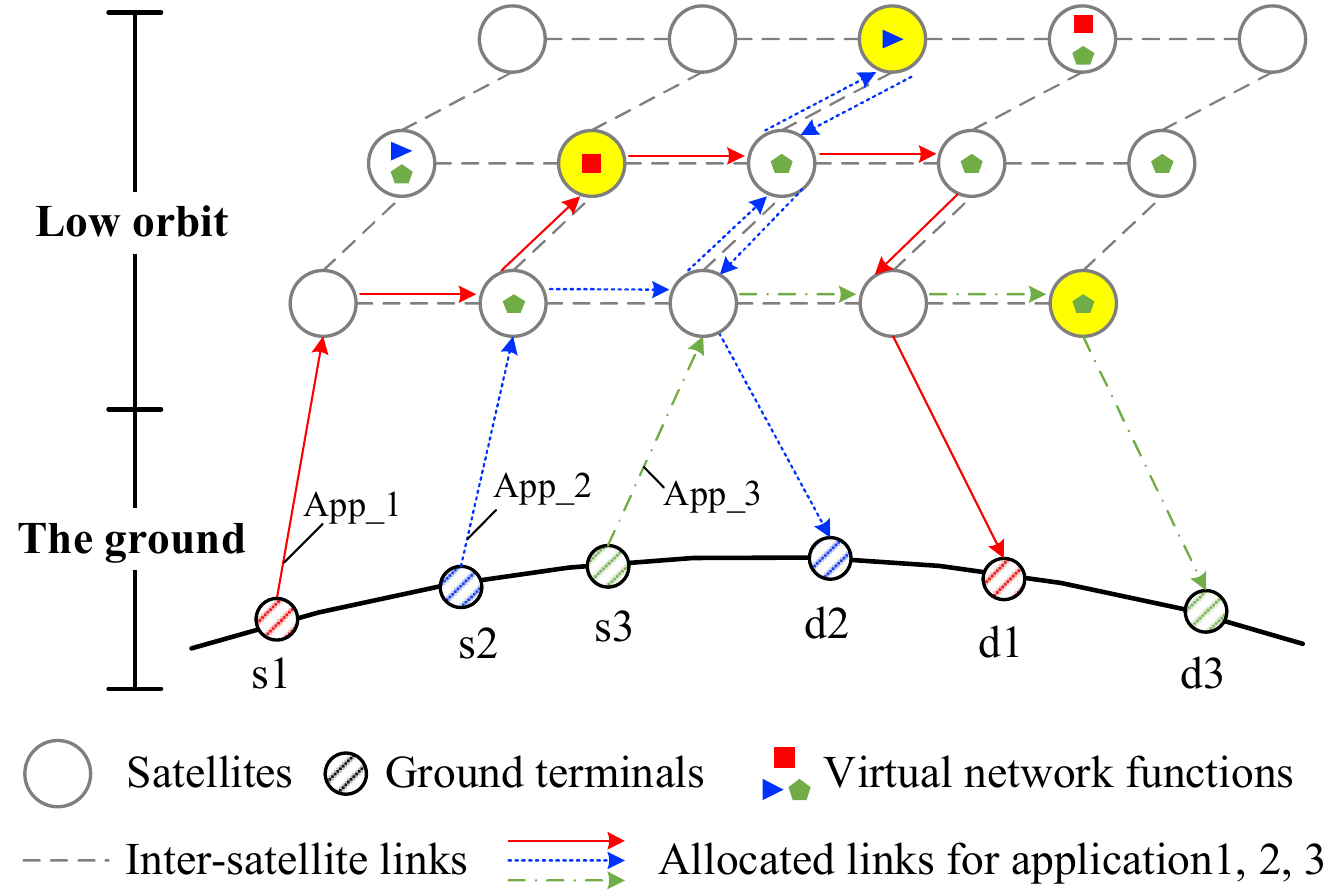}\
		\caption{An example of real time communications over satellite networks. }
	\end{figure}
	
	\textbf{An example.} 
	A scenario of an SN supporting three kinds of VFs is shown in \textcolor{black}{Figure 1}, where three VFs are marked by red square, blue triangle, green pentagon, respectively. There are three real time applications, and each of them has a pair of source and sink GT (e.g., \{s1, d1\}), Qos requirements and the specified function. The allocated paths are presented, and the selected satellites providing functions are highlighted in yellow. For \textit{App\_1} and \textit{App\_3}, simple paths bypassing a satellite deployed with the requested VF are allocated. For \textit{App\_2}, due to the scarcity of its specified VF (i.e., the blue triangle), a non-simple path with repeated nodes is provisioned.

    \subsection{Time-varying graph model}

In our system model, we divide the time horizon $\mathcal{T} = [0, T]$ into variable length time windows using the time division method in \cite{wang2022enhancing}. Each time window is denoted by $\tau = [t_s, t_e]$, where $t_s$ and $t_e$ are the start and end time of $\tau$, and $|\tau| = t_e - t_s $ is its length, $0 \leq t_s < t_e \leq T$. We split $\mathcal{T}$ in such a way that in each time window $\tau$, the link connectivities and their attributes (i.e., link delay and transmission capacity) remain unchanged. 

We use the snapshot graphs to describe the dynamic evolution of the topology in discrete time windows. 
	The snapshot in time interval $\tau = [t_{i-1}, t_i]$ is depicted as a \textcolor{black}{directed} graph $\mathcal{G}^{\tau} = (\mathcal{V}^{\tau}, \mathcal{L}^{\tau})$, where $\mathcal{V}^{\tau} = \mathcal{S} \cup \mathcal{O}$ is the node set including both the satellites and GTs, and  $\mathcal{L}^{\tau} = \{ (\mathbb{V}_i, \mathbb{V}_j) | \text{ node }\mathbb{V}_i \text{ is in the coverage of node } \mathbb{V}_j\}$ is the set of communication links (i.e., opportunities) including both the ISLs and ground satellite links (GSLs) during time interval $\tau$. 
	
	For each satellite $\mathbb{S}_i \in \mathcal{S}$, the maximum allowable numbers of calls of each function are collected into a set $\mathcal{W}_{\mathbb{S}_i} = \{w^{f_1}_{\mathbb{S}_i}, w^{f_2}_{\mathbb{S}_i}, ..., w^{f_N}_{\mathbb{S}_i}\}$, where $|\mathcal{W}_{\mathbb{S}_i}| = N$.
	Each link $(\mathbb{V}_i, \mathbb{V}_j)$ is characterized by its transmission rate $r^{\tau}_{(\mathbb{V}_i, \mathbb{V}_j)}$ (unit: Mbps) and its delay $D^{\tau}_{(\mathbb{V}_i, \mathbb{V}_j)}$ (unit: ms), where $D^{\tau}_{(\mathbb{V}_i, \mathbb{V}_j)}$ is the sum of the propagation delay, transmission delay and queuing delay. 
	\section{Problem Formulation}
	
	
	\subsection{Basic constrains for path selection}
	
		We define the binary variable ${x}^{\tau}_{\mathbb{V}_i, \mathbb{V}_j}$ for each link $(\mathbb{V}_i, \mathbb{V}_j) \in \mathcal{L}^{\tau}$, where ${{x}^{\tau}_{\mathbb{V}_i, \mathbb{V}_j} = 1}$ means the allocated path for the application $\mathcal{A}$ at time interval $\tau$ will pass through link $(\mathbb{V}_i, \mathbb{V}_j)$.
		Inherently, there are the following constraints. 
	\subsubsection{\text{Source node constraint}} Obviously, for the source GT, the path must come out from one of its adjacent links, namely, 
	\begin{equation}
	\sum_{\mathbb{V}_k \in \mathcal{V}^{\tau} - \{\mathbb{V}_s\}} {x}^{\tau}_{\mathbb{O}_s, \mathbb{V}_k} = 1. \label{c_source_node_constraints}
	\end{equation}
	\begin{equation}
	\sum_{\mathbb{V}_k \in \mathcal{V}^{\tau} - \{\mathbb{V}_s\}} {x}^{\tau}_{\mathbb{V}_k, \mathbb{O}_s} = 0. \label{c_source_node_remove_loop_constraints}
	\end{equation}
	\subsubsection{\text{Sink node constraint}} Similarly, for the destination GT, the path must enter one of its adjacent links, namely, 
	\begin{equation}
	\sum_{\mathbb{V}_k \in \mathcal{V}^{\tau} - \{\mathbb{O}_d\}} {x}^{\tau}_{\mathbb{V}_k, \mathbb{O}_d} = 1. \label{c_sink_node_constraints}
	\end{equation}
	\begin{equation}
	\sum_{\mathbb{V}_k \in \mathcal{V}^{\tau} - \{\mathbb{O}_d\}} {x}^{\tau}_{\mathbb{O}_d, \mathbb{V}_k} = 0. \label{c_sink_node_remove_loop_constraints}
	\end{equation}
	Note that the purpose of constraint (\ref{c_source_node_remove_loop_constraints}) and (\ref{c_sink_node_remove_loop_constraints}) is to prevent the loops of a path. However, sub-tours may still occur.

%
	
	\subsubsection{\text{Relay node constraint}} 
	For any relay node, a path can only pass through it no more than twice. That is, if the data is forwarded by it to reach the function node, the processed data will be transferred to the destination node through it again, or through other relay nodes. Thus, at most two of its incoming links can be included within the planned path, namely,  
	\begin{equation}
	\sum_{\mathbb{V}_k:(\mathbb{V}_k, \mathbb{V}_{\xi}) \in \mathcal{L}} {x}^{\tau}_{\mathbb{V}_k, \mathbb{V}_{\xi}} \leq 2, \forall \mathbb{V}_{\xi} \in \mathcal{V}^{\tau} -\{\mathbb{O}_s, \mathbb{O}_d\}. \label{c_incoming_node_constraints}
	\end{equation}
	And similarly, for its outgoing links, no more than two among them can be selected, namely,
	\begin{equation}
	\sum_{\mathbb{V}_k:(\mathbb{V}_{\xi},\mathbb{V}_k) \in \mathcal{L}} {x}^{\tau}_{\mathbb{V}_{\xi}, \mathbb{V}_k} \leq 2, \forall \mathbb{V}_{\xi} \in \mathcal{V}^{\tau} -\{\mathbb{O}_s, \mathbb{O}_d\}. \label{c_outgoing_node_constraints}
	\end{equation}
%
	
	\subsubsection{\text{Path integrity constraint}} If a path enters one incoming link of a forwarding node, it must also come out from one of its outgoing links. Thus, $ \forall \mathbb{V}_{\xi} \in \mathcal{V}^{\tau} -\{\mathbb{O}_s, \mathbb{O}_d\}$, 
	\begin{equation}
	\sum_{\mathbb{V}_k:(\mathbb{V}_k, \mathbb{V}_{\xi}) \in \mathcal{L}^{\tau}} {x}^{\tau}_{\mathbb{O}_k, \mathbb{V}_{\xi}}  = \sum_{\mathbb{V}_k:(\mathbb{V}_{\xi},\mathbb{V}_k) \in \mathcal{L}^{\tau}} {x}^{\tau}_{\mathbb{V}_{\xi}, \mathbb{V}_k}. \label{c_path_integrity_constraints}
	\end{equation}
	
	\subsubsection{\text{Path capacity constraint}} All of the transmission capacities of the links in the selected path must be larger than the application $\mathcal{A}$'s required capacity $C_a$, namely, 
	\begin{equation}
	 M \cdot (1-{x}^{\tau}_{\mathbb{V}_i, \mathbb{V}_j} ) + {x}^{\tau}_{\mathbb{V}_i, \mathbb{V}_j} \cdot {C}^{\tau}_{\mathbb{V}_i, \mathbb{V}_j} \geq C_a, \forall (\mathbb{V}_i, \mathbb{V}_j) \in \mathcal{L}^{\tau},  \label{c_path_capacity_constraints}
	\end{equation}
	where $M$ is a big constant commonly used in integer programming and can be set as the maximum communication capacity in SN. If ${x}^{\tau}_{\mathbb{V}_i, \mathbb{V}_j} = 1$, then $ M \cdot (1-{x}^{\tau}_{\mathbb{V}_i, \mathbb{V}_j}) = 0$, indicating the capacity of the selected link $ (\mathbb{V}_i, \mathbb{V}_j) $ must be greater than $C_a$. On the contrary, if ${x}^{\tau}_{\mathbb{V}_i, \mathbb{V}_j} = 0$, constraint (\ref{c_path_capacity_constraints}) becomes $M \geq C_a$ which automatically holds thus cancels the restriction on link $ (\mathbb{V}_i, \mathbb{V}_j) $'s capacity. 
	\subsubsection{\text{Path delay constraint}} The end-to-end delay of the selected path cannot exceed the required delay bound $D_a$. Intuitively, this constraint can be written as follows.
	\begin{equation}
	\sum_{(\mathbb{V}_i, \mathbb{V}_j) \in \mathcal{L}^{\tau}} {x}^{\tau}_{\mathbb{V}_i, \mathbb{V}_j} \cdot {D}^{\tau}_{\mathbb{V}_i, \mathbb{V}_j} \leq D_a. \label{c_path_delay_constraints}
	\end{equation}
	\subsubsection{\text{Virtual function constraint}} The selected path must contain at least one node providing the required function  $f_a$. Formally, such constraint can be written as follows.
	\begin{equation}
    \sum_{\mathbb{V}_\xi \in \mathcal{S}} [(\sum_{\mathbb{V}_k:(\mathbb{V}_k, \mathbb{V}_{\xi}) \in \mathcal{L}^{\tau}} {x}^{\tau}_{\mathbb{V}_k, \mathbb{V}_{\xi}} )\cdot w^{f_a}_{\mathbb{V}_{\xi}} ] \geq 1. \label{c_path_function_constraints}
	\end{equation}

	\subsection{Constraints for eliminating sub-tours}
	To eliminate possible \textit{sub-tours} in the path, we introduce the integer variable ${y}^{\tau}_{\mathbb{V}_i, \mathbb{V}_j}$ for each link $({\mathbb{V}_i, \mathbb{V}_j})$ to represent its link-order in the path, which can be regarded as the \textit{hop number} of the link $({\mathbb{V}_i, \mathbb{V}_j})$ at time slot $\tau$. If the link ${\mathbb{V}_i, \mathbb{V}_j}$ is included in the path, ${y}^{\tau}_{\mathbb{V}_i, \mathbb{V}_j} = h > 0$ indicates that the link $({\mathbb{V}_i, \mathbb{V}_j})$ is the $h-$hop of the path; otherwise, ${y}^{\tau}_{\mathbb{V}_i, \mathbb{V}_j} = 0$.  
	
	
	\subsubsection{\text{Binding variables}} 
	Firstly, if ${x}^{\tau}_{\mathbb{V}_i, \mathbb{V}_j} = 0$, then ${y}^{\tau}_{\mathbb{V}_i, \mathbb{V}_j} = 0$. Otherwise, $1 \leq {y}^{\tau}_{\mathbb{V}_i, \mathbb{V}_j} \leq \sum_{(\mathbb{V}_i, \mathbb{V}_j) \in \mathcal{L}_{\tau}}{x}^{\tau}_{\mathbb{V}_i, \mathbb{V}_j}$, where $\sum_{(\mathbb{V}_i, \mathbb{V}_j) \in \mathcal{L}_{\tau}}{x}^{\tau}_{\mathbb{V}_i, \mathbb{V}_j}$ is the maximum hop number of the planned path. Such a relationship can be expressed as follows.  
	\begin{equation}
	 {x}^{\tau}_{\mathbb{V}_i, \mathbb{V}_j} \leq {y}^{\tau}_{\mathbb{V}_i, \mathbb{V}_j} \leq {x}^{\tau}_{\mathbb{V}_i, \mathbb{V}_j} \cdot M. \label{c_bind_xy_constraints}
	\end{equation}
	\begin{equation}
	{y}^{\tau}_{\mathbb{V}_i, \mathbb{V}_j} \leq \sum_{(\mathbb{V}_i, \mathbb{V}_j) \in \mathcal{L}_{\tau}}{x}^{\tau}_{\mathbb{V}_i, \mathbb{V}_j}, \label{c_y_upper_bound_constraints}
	\end{equation}
	where $M$ is a big constant commonly used in logical constraints reformulation in integer programming, and can be set as $10^6$. It can be checked that if ${x}^{\tau}_{\mathbb{V}_i, \mathbb{V}_j} = 0$, then ${y}^{\tau}_{\mathbb{V}_i, \mathbb{V}_j} =0$; otherwise, ${y}^{\tau}_{\mathbb{V}_i, \mathbb{V}_j} <=M$ automatically holds and has no restriction on ${y}^{\tau}_{\mathbb{V}_i, \mathbb{V}_j}$.
Additionally, for source GT and destination GT, there is
\begin{equation}
\sum_{\mathbb{V}_k \in \mathcal{V}^{\tau} - \{\mathbb{V}_s\}} {y}^{\tau}_{\mathbb{O}_s, \mathbb{V}_k} = 1. \label{c_source_node_y0_constraints}
\end{equation}
\begin{equation}
\sum_{\mathbb{V}_k \in \mathcal{V}^{\tau} - \{\mathbb{O}_d\}} {y}^{\tau}_{\mathbb{V}_k, \mathbb{O}_d} = \sum_{(\mathbb{V}_i, \mathbb{V}_j) \in \mathcal{L}_{\tau}}{x}^{\tau}_{\mathbb{V}_i, \mathbb{V}_j}. \label{c_sink_node_y0_constraints}
\end{equation}

\subsubsection{\text{Sub-tour elimination constraints}} 
If the path passes through a forwarding node only once, the order of its outgoing link is the next hop of its incoming link, thus $ \sum_{\mathbb{V}_k:(\mathbb{V}_{\xi},\mathbb{V}_k) \in \mathcal{L}^{\tau}} {y}^{\tau}_{\mathbb{V}_{\xi}, \mathbb{V}_k} = \sum_{\mathbb{V}_k:(\mathbb{V}_k, \mathbb{V}_{\xi}) \in \mathcal{L}^{\tau}} {y}^{\tau}_{\mathbb{O}_k, \mathbb{V}_{\xi}} + 1$.
If the path passes through a forwarding node twice, the sum of the orders of its outgoing links is larger than the sum of the order of its incoming links by $2$, which is $ \sum_{\mathbb{V}_k:(\mathbb{V}_{\xi},\mathbb{V}_k) \in \mathcal{L}^{\tau}} {y}^{\tau}_{\mathbb{V}_{\xi}, \mathbb{V}_k} = \sum_{\mathbb{V}_k:(\mathbb{V}_k, \mathbb{V}_{\xi}) \in \mathcal{L}^{\tau}} {y}^{\tau}_{\mathbb{O}_k, \mathbb{V}_{\xi}} + 2$.  
Two different cases can be combined together as one equation, namely, $ \forall \mathbb{V}_{\xi} \in \mathcal{V}^{\tau} -\{\mathbb{O}_s, \mathbb{O}_d\}$, 
\begin{equation}
\sum_{\mathbb{V}_k:(\mathbb{V}_k, \mathbb{V}_{\xi}) \in \mathcal{L}^{\tau}}( {y}^{\tau}_{\mathbb{O}_k, \mathbb{V}_{\xi}} + {x}^{\tau}_{\mathbb{O}_k, \mathbb{V}_{\xi}}) = \sum_{\mathbb{V}_k:(\mathbb{V}_{\xi},\mathbb{V}_k) \in \mathcal{L}^{\tau}} {y}^{\tau}_{\mathbb{V}_{\xi}, \mathbb{V}_k}. \label{c_path_sub_tour_elim_constraints}
\end{equation}

	\subsection{Problem formulation}
	The objective is to minimize the end to end delay of the path satisfying the application $\mathcal{A}$'s requirements. 
	Therefore, the problem can be formulated as follows:
	\begin{equation*}
	\begin{split}
	\mathbf{P1: } \mathop  \text{min }  & \sum_{(\mathbb{V}_i, \mathbb{V}_j) \in \mathcal{L}^{\tau}} {x}^{\tau}_{\mathbb{V}_i, \mathbb{V}_j} \cdot {D}^{\tau}_{\mathbb{V}_i, \mathbb{V}_j}
	\\  \text{ s.t. }  & (\ref{c_source_node_constraints}) - (\ref{c_path_sub_tour_elim_constraints}).
	\end{split}
	\end{equation*}
	
	The problem \textbf{P1} is an integer linear programming (ILP) problem, as both the objective and constrains are linear functions. Such problem can be solved by commercial integer programming solvers, such as Gurobi \cite{gurobi}, using the classical branch and bound (B\&B) method. However, for even moderate scale SNs, searching for the optimal solution using the B\&B is still high, because the computational complexity of the B\&B is mainly related to the number of binary variables, the number of total constraints and the scale of the network. In particular, the total number of binary variables in \textbf{P1} is $|\mathcal{L}^{\tau}|$, and the total number of constraints is $H = 4 + (|\mathcal{V}^{\tau}|-2) \cdot 3 + 2 + |\mathcal{L}^{\tau}| = 3 |\mathcal{V}^{\tau}| + |\mathcal{L}^{\tau}| $. According to \cite{wolsey1999integer}, the B\&B method for solving \textbf{P1} has the worst case time complexity of $\mathcal{O}(|\mathcal{L}^{\tau}| \cdot H^{|\mathcal{L}^{\tau}|})$, which is exponential with the total number of links $|\mathcal{L}^{\tau}|$. Even for a medium scale \textbf{P1}, the running time can be in the order of minutes to hours, or even days for larger network sizes. 
	Therefore, it is necessary to exploit the special structure of \textbf{P1} and obtain more efficient methods.
	\section{The proposed graph-based algorithms}
Basically, solving \textbf{P1} is equal to finding the minimum delay $(\mathbb{O}_s-\mathbb{O}_d)$ path with capacity no less than $C_a$ and including at least one satellite deployed with function $f_a$.
In the following, firstly, we remove some capacity insufficient links and adopt the k-shortest path (KSP) algorithm\textcolor{black}{\cite{yen1971finding}} to solve it. After analyzing its drawbacks, an alternative algorithm is proposed by exploiting the special structures of the solution space.

\subsection{KSP-based VF constrained simple path algorithm}


The main idea is to iteratively find the $k$-th shortest delay paths from source GT $\mathbb{O}_s$ to destination GT $\mathbb{O}_d$ in the residual snapshot graph removing capacity-insufficient links, and stop until the required path is found or the $k$-th path violates the required path delay bound $D_a$. Firstly, all the links with capacity less than $C_a$ are removed to trim the solution space. 
Then, the searching procedure starts from the shortest delay path and each iteration a $k$-th shortest path $p_k$ is generated and checked.  If the new generated path $p_k$ satisfies the delay bound requirement, it will be checked whether it contains a satellite supporting function $f_a$. If a feasible path is found, the searching procedure terminates. Otherwise, it will stop until the delay of the $k$-th path exceeds the required delay $D_a$. 
The detailed KSP-based algorithm is shown in Algorithm 1.

\begin{algorithm}
	\caption{KSP-based VF constrained simple path algorithm}
	\label{array-sum}
	\hspace*{0.02in} {\bf Input:}
	$ \mathcal{G}^{\tau}=\{\mathcal{V}^{\tau}, \mathcal{L}^{\tau} \}$,
	and $\mathcal{A} = \{\mathbb{O}_s, \mathbb{O}_d, f_a, C_a, D_a \}$.  \\
	\hspace*{0.02in} {\bf Output:}
	The virtual function constrained simple path $p_{*}$.
	\begin{algorithmic}[1]
		\item  Initialize $\mathcal{G}^{\tau}_{C_a} \leftarrow \mathcal{G}^{\tau}$. 
		\item  \textbf{for} each link $(\mathbb{V}_i, \mathbb{V}_j) \in \mathcal{L}^{\tau}$ \textbf{do}
		\item  \ \ \ \  \textbf{if} $C^{\tau}_{(\mathbb{V}_i, \mathbb{V}_j)} < C_a $ \textbf{then}
		\item  \ \ \ \ \ \ \ \   \text{Remove} link $(\mathbb{V}_i, \mathbb{V}_j) $ from $\mathcal{G}^{\tau}_{C_a} $.
		\item $k=0$, $p_{*} = \varnothing $, $D_{p_0} = 0 $.
		\item \textbf{while} $D_{p_k} \leq D_a $ \textbf{do}
		\item 	\hspace*{0.02in} \hspace*{0.02in}
		$k \longleftarrow k+1$
		\item 	\hspace*{0.02in} \hspace*{0.02in}
		Find the $k$-th shortest path $p_k$ from $\mathbb{O}_s$ to $\mathbb{O}_d$ in $\mathcal{G}^{\tau}_{C_a}$.
		\item 	\hspace*{0.02in} \hspace*{0.02in}
		Calculate the delay of the $k$-th path $p_k$ as $D_{p_k}$.
		\item  \hspace*{0.02in} \hspace*{0.02in} \textbf{if} $\exists \mathbb{S}_{\xi} \in p_k, w^{f_a}_{\mathbb{S}_{\xi}}\geq 1 $ \textbf{then}
		\item  \hspace*{0.02in} \hspace*{0.02in} 	\hspace*{0.02in} \hspace*{0.02in} $p_{*} = p_k$, break. 
		\item \textbf{return} $p_{*}$.
	\end{algorithmic}
\end{algorithm}

Obviously, the running time of \text{Algorithm 1} depends on how many paths within delay bound $D_a$ need be iterated before a path containing a satellite deployed with function $f_a$ is found.
The drawbacks of Algorithm 1 are summarized as follows:

\begin{itemize}
	
	\item  \textit{Sub-optimality:} Since the k-shortest path algorithm computes simple paths with no repeating nodes, the optimal path which have repeated satellites can not be obtained by this algorithm. Even worse, such a drawback may cause applications to be rejected though there is more than one feasible non-simple path with repeated nodes. 
	
	\item \textit{Instability:} When the delay bound $D_a$ is small, the algorithm will stop quickly since there are fewer paths to be checked. However, when $D_a$ is large, it may take a very long time for the algorithm to iterate paths. Therefore, the running time of the algorithm is related to both the value of $D_a$ and the number of function enabled satellites, which makes the algorithm unstable and pseudo-polynomial.  
	
	
	\item  \textit{Non Scalability: } Although a single path can be found in polynomial time, e.g., $\mathcal{O}(|\mathcal{L}^{\tau}| + |\mathcal{V}^{\tau}|\log|\mathcal{V}^{\tau}|) $ by using the Dijkstra's algorithm, the number of infeasible paths in a densely connected graph with thousands of nodes can be extremely large, i.e., $\mathcal{O}(|\mathcal{V}|!)$ in a complete graph of order $|\mathcal{V}^{\tau}|$. For large scale SNs, the KSP-based method will be very time consuming. 
\end{itemize}   


\textit{Complexity analysis:} 
In fact, based on the KSP algorithm proposed by \cite{yen1971finding}, finding $K$ number of simple paths requires $O(K|\mathcal{V}^{\tau}|^3)$ operations. However, in the worst case, given a complete graph of order $|\mathcal{V}^{\tau}|$ with no function enabled satellite and $D_a = +\infty$, the computational complexity of Algorithm 1 is $O(|\mathcal{V}^{\tau}|! \cdot |\mathcal{V}^{\tau}|^3)$, which is prohibitively time consuming and can not be deployed in large scale SNs.

\subsection{The proposed VF-aware shortest path algorithm}

	In practice, both the ILP-based method and KSP-based method have drawbacks, preventing these two solvers from being applicable in large-scale SNs. 
	To address this bottleneck, we propose an alternative method by bidirectional path seeking from source and destination to the functional satellites (satellites can providing required functions), such that the problem can be solved optimally by running two times of Dijkstra's algorithm combined with a node selection procedure. 

	Instead of directly searching for candidate paths, we collect the functional satellites during the link filtering operation (same as Algorithm 1) before calculating the path. 
	As a result, the path seeking problem of \textbf{P1} can be decomposed into two sub-problems: 1) Seeking the shortest path from source GT to every functional satellites; 2) Seeking the shortest path from every functional satellites to destination GT. 
	Since there can be multiple functional satellites, the complexity of path calculation is too high. 
	Fortunately, the Dijkstra's algorithm has the special property that the computation of one single-source single-sink shortest path can obtain all the shortest paths from the single-source to all other destinations. In other words, the complexity of single-source single-sink shortest path is the same as that of the single-source multi-sink shortest path. 
	Therefore, we view the functional satellites as multiple sinks, and run two times of Dijkstra's algorithm, one in the forward direction from source GT to functional satellites, and another in the reverse direction from the destination GT to all functional satellites by reversing all links of the original snapshot graph. Finally, the VF-aware shortest path can be found by joining the partial paths, which is the optimal solution of \textbf{P1}.

\begin{algorithm}
	\caption{The VF-aware shortest path (VFSP) algorithm}
	\label{array-sum}
	\hspace*{0.02in} {\bf Input:}
	$\mathcal{G}^{\tau}=\{\mathcal{V}^{\tau}, \mathcal{L}^{\tau} \}$, and $\mathcal{A} = \{\mathbb{O}_s, \mathbb{O}_d, f_a, C_a, D_a \}$.\\
	\hspace*{0.02in} {\bf Output:} The virtual function constrained shortest path $p^*$. 
	\begin{algorithmic}[1]
		\item  Initialize $\mathcal{G}^{\tau}_{C_a} = \varnothing$, $\mathcal{V}^{\tau}_{f_a} = \emptyset$, $p_{*} = \varnothing$, and $D_{p_{*}} = D_a + 1 $. 
		\item  \textbf{for} each link $(\mathbb{V}_i, \mathbb{V}_j) \in \mathcal{L}^{\tau}$ \textbf{do}
		\item  \ \ \ \  \textbf{if} $C^{\tau}_{(\mathbb{V}_i, \mathbb{V}_j)} \geq  C_a $ \textbf{then}
		\item  \ \ \ \ \ \ \ \   \text{Add} link $(\mathbb{V}_i, \mathbb{V}_j) $ to $\mathcal{G}^{\tau}_{C_a} $.
		\item  \ \ \ \ \ \ \ \  \textbf{if} $w^{f_a}_{\mathbb{V}_{i}}\geq 1 $ \textbf{then}
		\item  \ \ \ \ \ \ \ \  \ \ \ \   $\mathcal{V}_{f_a} \leftarrow \mathcal{V}_{f_a} \cup \{\mathbb{V}_i\}$.
		\item  \ \ \ \  \ \ \ \  \textbf{if} $w^{f_a}_{\mathbb{V}_{j}}\geq 1 $ \textbf{then}
		\item  \ \ \ \ \ \ \ \ \ \ \ \   $\mathcal{V}_{f_a} \leftarrow \mathcal{V}_{f_a} \cup \{\mathbb{V}_j\}$.
		\item  \textbf{if} $\mathcal{V}_{f_a} = \emptyset $ \textbf{then}
		\item  \ \ \ \  \textbf{return} -1. // No feasible path. 
		\item  Reverse all links in $\mathcal{G}^{\tau}_{C_a}$ to build a reversed graph $\bar{\mathcal{G}}^{\tau}_{C_a}$.
		\item  Calculate shortest paths from  $\mathbb{O}_s$ to all other nodes in $\mathcal{G}^{\tau}_{C_a}$.
		\item  Calculate shortest paths from  $\mathbb{O}_d$ to all other nodes in $\bar{\mathcal{G}}^{\tau}_{C_a}$.
		\item  \textbf{for} each satellite $\mathbb{S}_i \in \mathcal{V}^{\tau}_{f_a} $ \textbf{do}
		\item  \ \ \ \  \textbf{if} $D_{p_{(\mathcal{O}_s, \mathbb{S}_i)}} + D_{p_{(\mathcal{O}_d, \mathbb{S}_i)}} \leq D_{p_{*}}$ \textbf{then}
		\item  \ \ \ \ \ \ \ \   $D_{p_{*}} = D_{p_{(\mathcal{O}_s, \mathbb{S}_i)}} + D_{p_{(\mathcal{O}_d, \mathbb{S}_i)}}$, $p_{*} =p_{(\mathcal{O}_s, \mathbb{S}_i)} + \bar{p}_{(\mathcal{O}_d, \mathbb{S}_i)} $.
		\item  {\bf return}  $p_*$.
	\end{algorithmic}
\end{algorithm}
\textit{Complexity analysis:}
Based on the VFSP, finding all shortest paths and functional satellites requires $O(|\mathcal{L}^{\tau}|)$ operations. And the computational complexity for Dijkstra's algorithm for computing one source multiple sink shortest path is $\mathcal{O}(|\mathcal{L}^{\tau}|+|\mathcal{V}^{\tau}| \log{|\mathcal{V}^{\tau}|})$. Therefore, the worst case computational complexity of Algorithm 2 is $\mathcal{O}(3|\mathcal{L}^{\tau}| + (2\log{|\mathcal{V}^{\tau}|}+1) |\mathcal{V}^{\tau}|)$, which is polynomial with the increase of network size. 

\section{Evaluation}
\subsection{Simulation setup}
We conduct the simulation based on the starlink constellation, which is the largest LEO satellite system by far.
 Specifically, we randomly choose $100$-$2300$ satellites from 2694 active starlink satellites in standard object database of systems tool kit (STK),
a third-party software that updates continuously to simulate the movement of real-world satellites. 
The ground terminals are randomly distributed and located in around Xi'an ( $34.27^{\circ} N, 108.93^{\circ} E$ ), Beijing ($40^{\circ} N, 116^{\circ} E$), Sanya ( $18^{\circ} N, 109.5^{\circ} E$) and Kashi ($39.5^{\circ} N, 76^{\circ} E$). 
The contact plans including communication opportunities of each pair of satellites in the studied network are obtained by using \textit{compute access} function in STK. 
The transmission rate of both ISLs and GSLs is uniformly selected from $[300,350]$ Mbps as in \cite{fu2020remote}.
The link delays of both ISLs and GSLs are in the range of $[5, 15]$ ms as in \cite{chen2022robust}.

The simulation time horizon $T$ is from \text{2022-8-9 04:00:00} to \text{2022-8-10 04:00:00}, and is split into multiple time windows by the time division mechanism in \cite{wang2022enhancing}. 
All the three algorithms are implemented using Python programming language. 

\subsection{Simulation results and analysis}
\subsubsection{Running time performance with increasing satellite network sizes}
Fix the deployed function number $|\mathcal{F}|=3$ and the percentage of function enabled satellites as $10\%$, we consider 5000 random applications. Each application involves a pair of GTs, and requires the transmission capacity and the end to end delay randomly selected from $[5, 100]$ Mbps, and $[20, 150]$ ms, respectively.
\textcolor{black}{Figure 2} reports the running time of three different algorithms with the number of satellites varying from $100$ to $2300$, with a step-wise increase of $100$.  
As the network size increases, the average running time of all the three algorithms increases. 
However, both the graph-based methods (KSP, VFSP) are significantly faster than the ILP method. 
This can be explained as follows. For the ILP-based method, as the satellite number becomes large, more transmission edges are added into the snapshot graph, resulting in increased number of decision variables and exponential growing running times. Similarly, the increased transmission opportunities also introduce more infeasible paths in the snapshot graph for the KSP-based method to explore, thus the number of iterations increases. Since the computation complexity of the proposed VFSP mainly depends on the function enabled satellite numbers in a network, its running time increases linearly with the increase of satellite numbers.

\begin{figure}[t!]
	\centering
		\includegraphics[width=60mm]{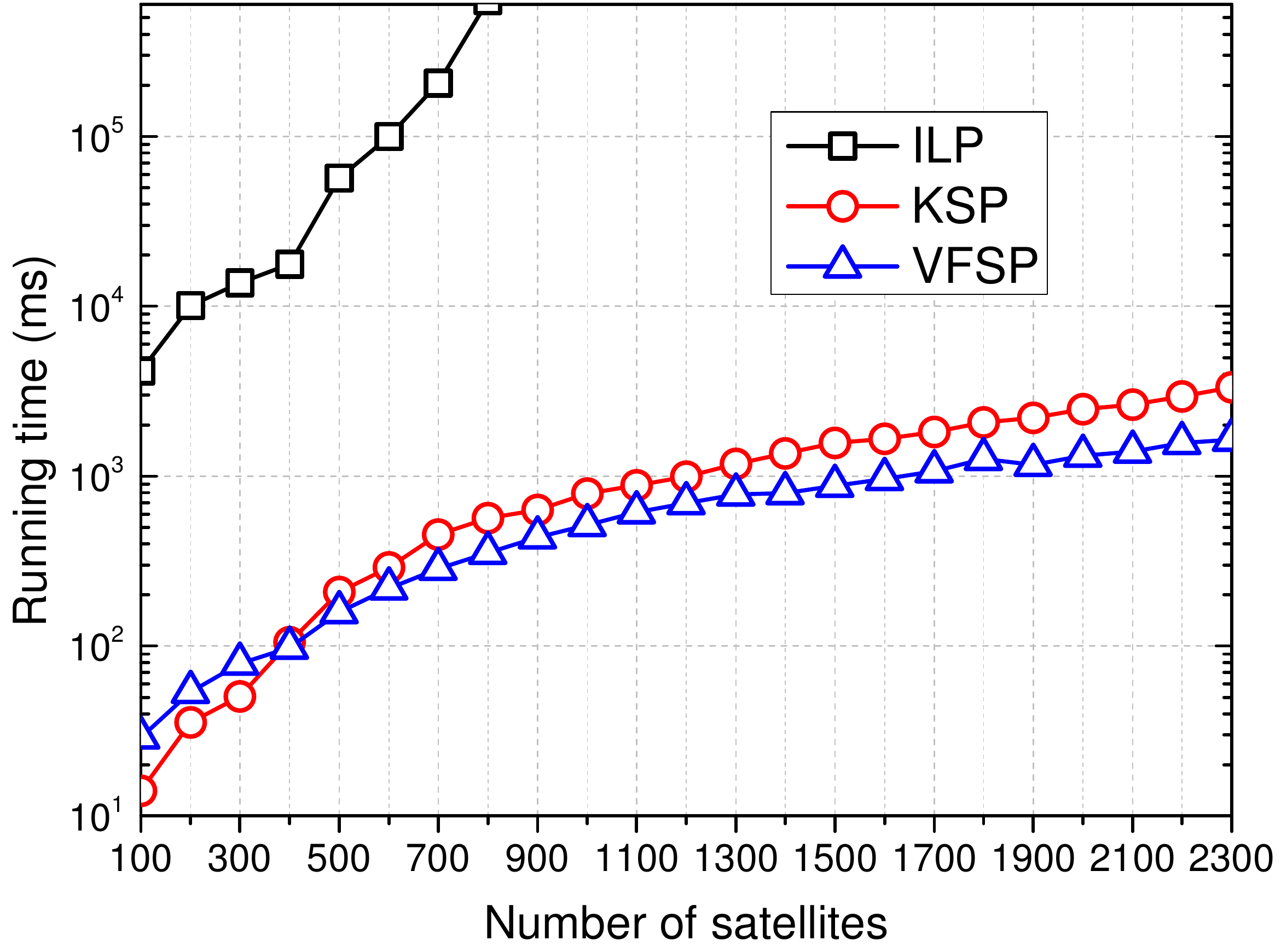}\\
	\caption{Running times versus different satellite numbers. }
\end{figure}


\begin{figure}[t!]
	\centering
			\includegraphics[width=60mm]{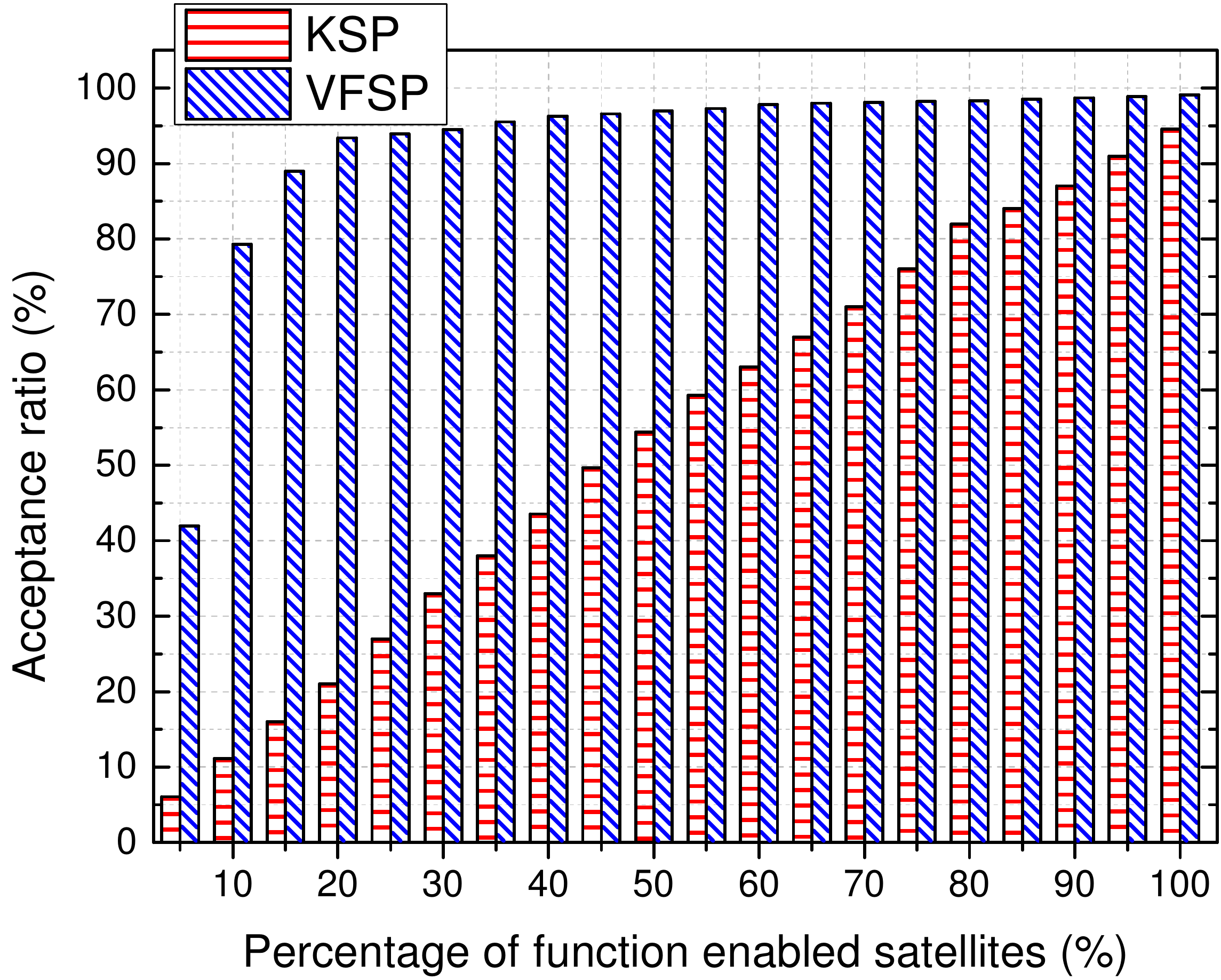}\\
	\caption{Acceptance ratios versus function enabled satellite numbers. }
\end{figure}


\subsubsection{Network performance with increasing function enabled satellites}
Fix the satellite number as $1000$, we vary the percentage of function enabled satellites to investigate the impact of the function deployment ratio on network performance. In Figure 3, the acceptance ratio versus the percentage of function enabled satellites is plotted from $5\%-100\%$ with an increase step of $5\%$, where $5\%$ means that $50$ out of $1000$ satellites are deployed with virtual functions. For comparison, each function enabled satellite accommodates at most one application. As seen from Figure 3, more function enabled satellites can bring higher acceptance ratio for both the KSP and VFSP. It is as expected that the acceptance ratio of the VFSP is higher than that of the KSP, since the VFSP can explore more non-simple paths to accommodate applications while the KSP can not. Moreover, when the percentage of function enabled satellites is less than $20\%$, the acceptance ratio grows faster since the function resources are the main bottleneck of the network. 

Figure 4 and Figure 5 plot the average delay and average number of hops of the paths with the increase of the percentage of function enabled satellites, respectively. The path delay of the KSP is smaller than that of the VFSP, which is expected since applications with no feasible simple paths are rejected by the KSP.
However, such applications can be fulfilled by the VFSP with non-simple paths with repeated nodes thus longer delays. The trend in Figure 5 is similar to that of Figure 4, as smaller end to end delays correspond to fewer hop numbers. However, the lower delay and fewer hops performance of the KSP method is at the cost of reduced acceptance ratios, since a large proportion of feasible applications can be rejected.


\begin{figure}[t!]
	\centering
	\includegraphics[width=60mm]{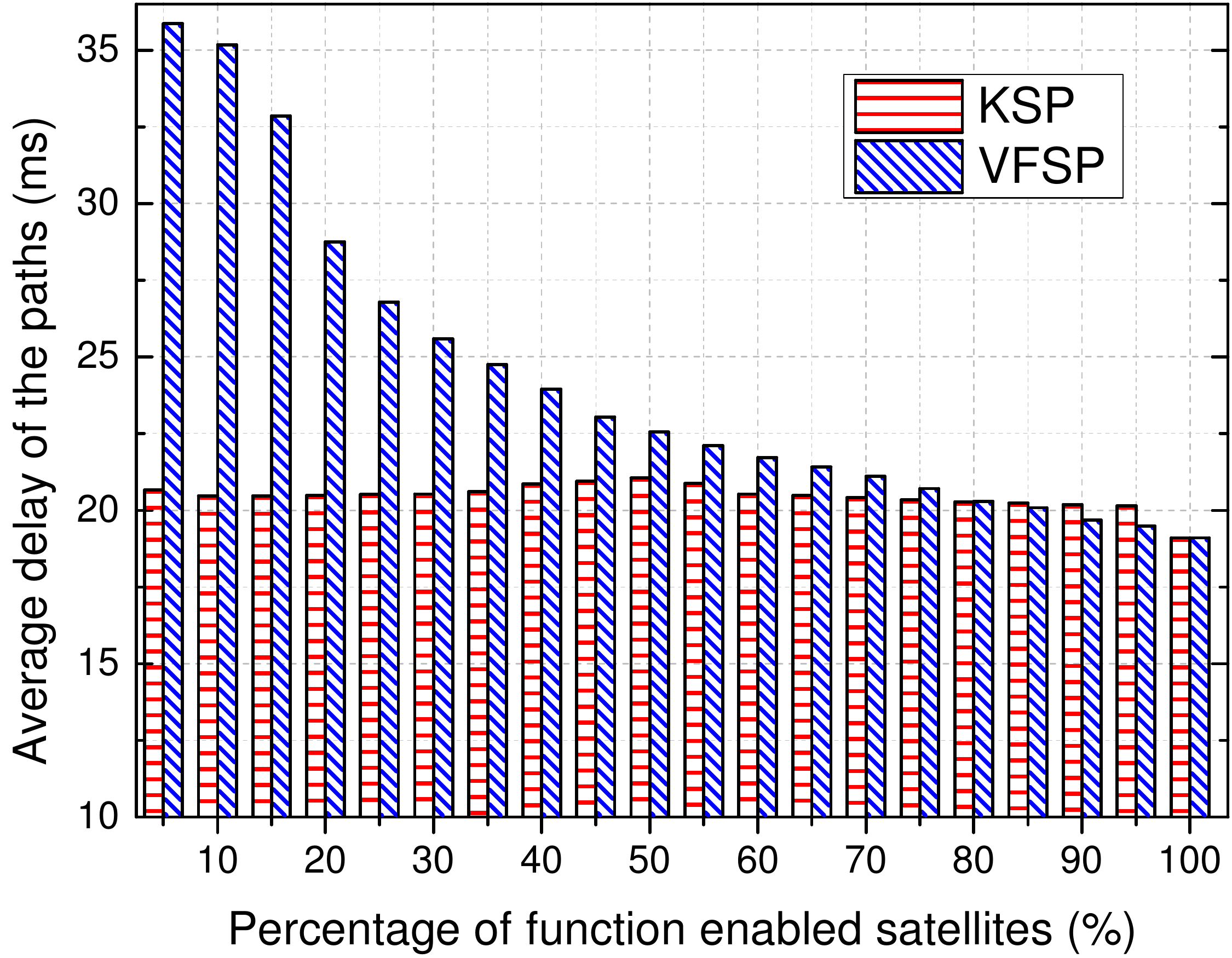}\\
	\caption{Path delays versus function enabled satellites. }
\end{figure}

\begin{figure}[t!]
	\centering
	\includegraphics[width=60mm]{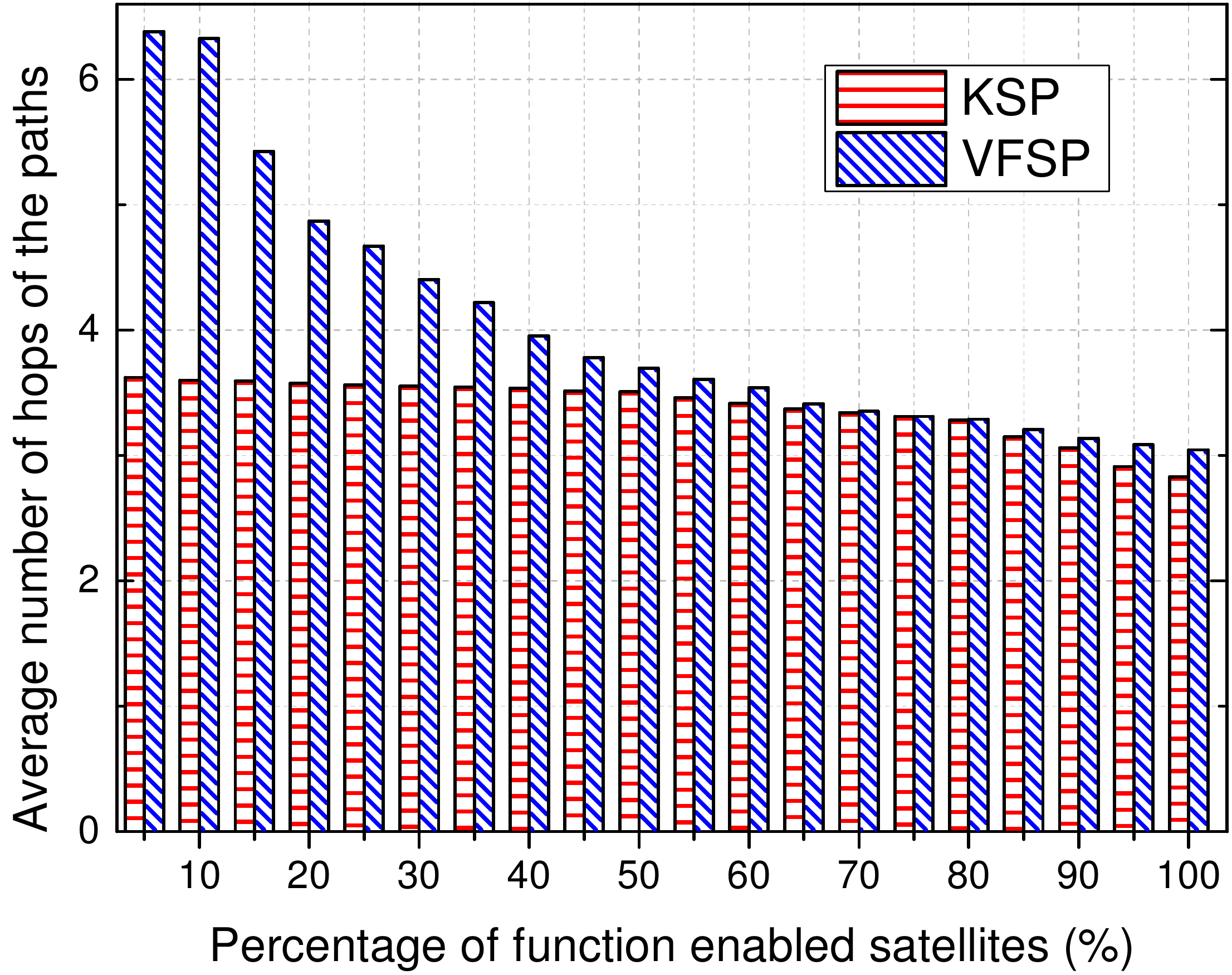}\\
	\caption{Number of hops versus different satellite numbers. }
\end{figure}

	\section{Conclusion}
	

In this work, we investigate the routing strategy for real time applications in large-scale SNs with VFs. We identify that such a routing problem can be formulated as an integer linear programming problem, which incurs exponential complexity by using the branch and bound method. To solve the problem efficiently, two alternative graph-based algorithms from different standpoints are proposed, analyzed and compared, where the latter one can obtain the optimal solution in polynomial time with a low computation complexity and more stable performance in large scale networks. Simulations conducted on starlink constellation with thousands of satellites verify the performance of the proposed algorithms. 

	\ifCLASSOPTIONcaptionsoff
	\newpage
	\fi

	
	
	%
	
	\bibliographystyle{IEEEtran}
	\bibliography{reference}

\begin{thebibliography}{10}
\providecommand{\url}[1]{#1}
\csname url@samestyle\endcsname
\providecommand{\newblock}{\relax}
\providecommand{\bibinfo}[2]{#2}
\providecommand{\BIBentrySTDinterwordspacing}{\spaceskip=0pt\relax}
\providecommand{\BIBentryALTinterwordstretchfactor}{4}
\providecommand{\BIBentryALTinterwordspacing}{\spaceskip=\fontdimen2\font plus
\BIBentryALTinterwordstretchfactor\fontdimen3\font minus
  \fontdimen4\font\relax}
\providecommand{\BIBforeignlanguage}[2]{{%
\expandafter\ifx\csname l@#1\endcsname\relax
\typeout{** WARNING: IEEEtran.bst: No hyphenation pattern has been}%
\typeout{** loaded for the language `#1'. Using the pattern for}%
\typeout{** the default language instead.}%
\else
\language=\csname l@#1\endcsname
\fi
#2}}
\providecommand{\BIBdecl}{\relax}
\BIBdecl

\bibitem{ITU2021}
ITU, ``Facts and figures 2021: 2.9 billion people still offline,''
  \url{https://www.itu.int/hub/2021/11/facts-and-figures-2021-2-9-billion-people-still-offline/},
  2021.

\bibitem{lai2022spacertc}
Z.~Lai, W.~Liu, Q.~Wu, H.~Li, J.~Xu, and J.~Wu, ``Space{RTC}: Unleashing the
  low-latency potential of mega-constellations for real-time communications,''
  in \emph{Proc. IEEE Conf. on Comput. Commun. (INFOCOM)}, 2022, pp.
  1339--1348.

\bibitem{sheng2017toward}
M.~Sheng, Y.~Wang, J.~Li, R.~Liu, D.~Zhou, and L.~He, ``Toward a flexible and
  reconfigurable broadband satellite network: Resource management architecture
  and strategies,'' \emph{IEEE Wireless Commun.}, vol.~24, no.~4, pp. 127--133,
  2017.

\bibitem{xu2018software}
S.~Xu, X.-W. Wang, and M.~Huang, ``Software-defined next-generation satellite
  networks: Architecture, challenges, and solutions,'' \emph{IEEE Access},
  vol.~6, pp. 4027--4041, 2018.

\bibitem{bertaux2015software}
L.~Bertaux, S.~Medjiah, P.~Berthou, S.~Abdellatif, A.~Hakiri, P.~Gelard,
  F.~Planchou, and M.~Bruyere, ``Software defined networking and virtualization
  for broadband satellite networks,'' \emph{IEEE Commun. Magazine}, vol.~53,
  no.~3, pp. 54--60, 2015.

\bibitem{zhou2019bidirectional}
S.~Zhou, G.~Wang, S.~Zhang, Z.~Niu, and X.~S. Shen, ``Bidirectional mission
  offloading for agile space-air-ground integrated networks,'' \emph{IEEE
  Wireless Commun.}, vol.~26, no.~2, pp. 38--45, 2019.

\bibitem{wang2020sfc}
G.~Wang, S.~Zhou, S.~Zhang, Z.~Niu, and X.~Shen, ``{SFC}-based service
  provisioning for reconfigurable space-air-ground integrated networks,''
  \emph{IEEE J. Sel. Areas Commun.}, vol.~38, no.~7, pp. 1478--1489, 2020.

\bibitem{jia2021vnf}
Z.~Jia, M.~Sheng, J.~Li, D.~Zhou, and Z.~Han, ``{VNF}-based service provision
  in software defined leo satellite networks,'' \emph{IEEE Trans. Wireless
  Commun.}, vol.~20, no.~9, pp. 6139--6153, 2021.

\bibitem{yang2021maximum}
H.~Yang, W.~Liu, H.~Li, and J.~Li, ``Maximum flow routing strategy for space
  information network with service function constraints,'' \emph{IEEE Trans.
  Wireless Commun.}, vol.~21, no.~5, pp. 2909--2923, 2021.

\bibitem{wang2022enhancing}
P.~Wang, H.~Li, B.~Chen, and S.~Zhang, ``Enhancing earth observation throughput
  using inter-satellite communication,'' \emph{IEEE Trans. Wireless Commun.},
  2022.

\bibitem{gurobi}
\BIBentryALTinterwordspacing
L.~Gurobi~Optimization, ``Gurobi optimizer reference manual,'' 2021. [Online].
  Available: \url{http://www.gurobi.com}
\BIBentrySTDinterwordspacing

\bibitem{wolsey1999integer}
L.~A. Wolsey and G.~L. Nemhauser, \emph{Integer and combinatorial
  optimization}.\hskip 1em plus 0.5em minus 0.4em\relax John Wiley \& Sons,
  1999, vol.~55.

\bibitem{yen1971finding}
J.~Y. Yen, ``Finding the k shortest loopless paths in a network,''
  \emph{management Science}, vol.~17, no.~11, pp. 712--716, 1971.

\bibitem{fu2020remote}
W.~Fu, J.~Ma, P.~Chen, and F.~Chen, ``Remote sensing satellites for digital
  earth,'' in \emph{Manual of digital earth}.\hskip 1em plus 0.5em minus
  0.4em\relax Springer, 2020, pp. 55--123.

\bibitem{chen2022robust}
Q.~Chen, W.~Meng, S.~Han, C.~Li, and H.-H. Chen, ``Robust task scheduling for
  delay-aware iot applications in civil aircraft-augmented sagin,'' \emph{IEEE
  Trans. Commun.}, vol.~70, no.~8, pp. 5368--5385, 2022.

\end{thebibliography}

\end{document}